\documentclass[twocolumn,english,preprintnumbers,amsmath,amssymb,floatfix]{revtex4}
\usepackage[T1]{fontenc}
\usepackage[latin9]{inputenc}
\usepackage{color}
\usepackage{array}
\usepackage{amstext}
\usepackage{graphicx}
\usepackage{esint}
\usepackage{epsfig}
\usepackage{graphics}
\usepackage{latexsym}
\usepackage{amsmath}
\usepackage{amssymb}
\usepackage{rotating}
\usepackage{subfigure}
\usepackage{bm}
\usepackage{ulem}

\usepackage[colorlinks = true,
linkcolor = magenta,
urlcolor  = blue,
citecolor = red,
anchorcolor = blue]{hyperref}

\makeatletter

\@ifundefined{textcolor}{}
{%
 \definecolor{BLACK}{gray}{0}
 \definecolor{WHITE}{gray}{1}
 \definecolor{RED}{rgb}{1,0,0}
 \definecolor{GREEN}{rgb}{0,1,0}
 \definecolor{BLUE}{rgb}{0,0,1}
 \definecolor{CYAN}{cmyk}{1,0,0,0}
 \definecolor{MAGENTA}{cmyk}{0,1,0,0}
 \definecolor{YELLOW}{cmyk}{0,0,1,0}
 }

\@ifundefined{definecolor}
 {\usepackage{color}}{}
\@ifundefined{definecolor}
 {\usepackage{color}}{}
\makeatother

\makeatother

\usepackage{babel}

\begin{document}

\title{Hadron production through Higgs decay at next-to-leading order in the general-mass variable-flavor-number scheme
	 
}

\author{S. Mohammad Moosavi Nejad }

\email{mmoosavi@yazd.ac.ir}

\affiliation{$^{(a)}$Department of Physics, Yazd University, P.O. Box
89195-741, Yazd, Iran}

\date{\today}

\begin{abstract}

It is known that  about  $60\%$ of all Higgses  produced at the CERN-LHC decay into a pair of bottom quarks. Bottoms quickly hadronize, in most cases, into bottom-flavored (B) hadrons before they decay. Therefore, the study of   scaled-energy  distribution of B-mesons in the decay process $H\to B+Jets$ can be considered as a channel to search for the Higgs characteristics. In all previous studies,  authors have ignored  the mass effect of b-quarks as well as B-mesons by working in the massless scheme. In this work we, for the first time, study the mass effect of b-quarks as well as produced mesons on the  scaled-energy ($x_B$) distribution of B-mesons  by working in the massive scheme or  general-mass variable-flavor-number scheme (GM-VFNs). We find that the meson  mass is responsible for a significant enhancement of partial decay width in the low-$x_B$ region while the b-quark mass  leads  to an enhancement of the partial decay rate in the peak region and above.

\end{abstract}


\maketitle

\section{Introduction}

Observation of the Higgs boson by the CERN-ATLAS and CMS Collaborations in 2012 completed
the standard model (SM) of strong and electroweak interactions provided it is compatible with the SM Higgs boson. During the last decades a lot of knowledge has  been learned about this key particle and its discovery was an important breakthrough in our understanding of fundamental interactions. Nevertheless, a basic question  is still  whether the characteristics of observed particle is in complete consistency with that predicted by the SM. Consistencies with the SM Higgs boson needs to be examined in more detail by measuring its couplings to other SM particles. These are pursued by deriving the related couplings from the study of Higgs boson decays and its production rates at the LHC  and will be continued in the future runs.
When studying the Higgs boson,  particle physicists  have to look for its decay products in their detectors, as the Higgs  decays very quickly after it is produced.
Note that Higgs bosons can decay to different particles but not all final states are equally likely. For instance, around $60\%$ of all Higgses produced at the LHC decay into a pair of bottom quarks ($H\to b\bar{b}$), while only $3\%$ decay into a pair of $Z$ bosons \cite{CMS:2018zzl}.  
In recent years, the LHC has achieved great success in discovering the Higgs couplings  to charged fermions of  third generation as well as heavy vector bosons  \cite{CMS:2017zyp,CMS:2018nsn,ATLAS:2018kot}. Note that,  in the SM the Higgs couplings  to the first and second generation fermions are so  weak and very difficult to be measured directly. Nevertheless,  due to  complicated hadronic backgrounds and the limited Higgs events the precision of these measurements is restricted.\\
At the CERN-LHC, we expect to have more data  collected during Run~3  than the first two runs combined. Moreover, this machine  is planned to upgrade to the high-energy and high-luminosity LHC  after Run~3 so it is anticipated to have about $1.6\times 10^8$  Higgs events at the energy $\sqrt{s}=14$~TeV with an integrated luminosity of $3$ ab$^{-1}$ and also  $2.2\times 10^9$  Higgs events at the energy $\sqrt{s}=27$~TeV with integrated luminosity of $15$ ab$^{-1}$ \cite{Cepeda:2019klc}. Beside that, several new high energy colliders such as  the Circular Electron-Positron Collider (CEPC) \cite{CEPCStudyGroup:2018ghi},   the $e^+e^-$ Future Circular Collider (FCC-ee) \cite{FCC:2018evy}, the International Linear Collider (ILC) \cite{ILC:2013jhg} and the Muon Collider \cite{Black:2022cth} are under consideration so their suitable   advantage is to provide a clean environment  for the exact measurements of Higgs properties. These future generation colliders allow particle physicists to measure some rare decays of  Higgses such as their decays to heavy quarkonia \cite{Bodwin:2013gca,Liao:2018nab,Pan:2022nxc,CDF:2015eag}. These are necessary to determine the magnitude of Yukawa couplings of  Higgs to heavy quark flavors. Decays to heavy quarkonia have distinguished signals which can be detected at the high-energy  colliders. \\
As was mentioned, at the LHC it is observed that  around $60\%$ of Higgs bosons decay  into a pair of bottom quarks so, in the following, bottom quarks hadronize before they decay. In most cases, bottom quarks hadronize into bottom-flavored mesons $B$, however other hadron species are possible to be produced.
Therefore, investigating the energy spectrum of B-mesons produced through Higgs boson decay in the process $H\to b\bar{b}\to B+Jets$ would  be proposed as a channel to look for the Higgs characteristics.  Moreover, this proposed channel can also be used to study the production mechanism of B-mesons, which is described by the nonperturbative fragmentation functions (non-pFFs). With these motivations, in many papers the energy spectrum of B-mesons have been studied  by evaluating  the quantity $d\Gamma(H\to BX)/dx_B$ where $x_B(=2p_B\cdot p_H/p_H^2=2E_B/m_H)$ is the scaled-energy of B-meson in the Higgs boson rest frame, see for example Refs.~\cite{Corcella:2004xv,Zheng:2023atb,Baradaran:2025khn}.
In all previous works, for simplicity, the mass effects of bottom quarks  as well as produced mesons have been ignored. This choice is known as the massless scheme or zero-mass variable-flavor-number scheme (ZM-VFNs)  where the assumptions $m_b=0$ and $m_B=0$ are  set from the beginning.
Here, we revisit bottom-flavored hadron production from the SM Higgs decay 
by working  in the general-mass variable-flavor-number scheme (GM-VFNs)  which provides an ideal theoretical framework to taking into account the effects of quark masses.
Being manifestly based on QCD-improved parton model factorization theorem (or  Collin's hard-scattering factorization theorem) \cite{collins}
 appropriate  for massive quarks, this 
factorization scheme allows one  to resum the large logarithms 
in $m_b$, i.e. $\ln (m_H^2/m_b^2)$, to retain the finite-$m_b$ effects and to preserve the universality of the non-perturbative fragmentation functions, 
whose scaling violations remain to be subject to DGLAP evolution equations \cite{dglap}.
In this way, the GM-VFNs  combines the virtues of  fixed-flavor-number scheme (FFNs) and the ZM-VFN scheme  and, at the same time, this scheme avoids their flaws. In fact, the GM-VFNs  is a  suitable  tool for global analyses of experimental 
data on the inclusive production of heavy hadrons, allowing us  to transfer 
nonperturbative information on the hadronization of partons (quarks or gluon)
from one type of experiment to another and from one 
energy scale $\mu_F$ to another one, without the restriction $\mu_F\gg m_b$ which is necessary for the ZM-VFNs (more detail about the GM-VFNs can be found in Refs.~\cite{Kneesch:2007ey,Yarahmadi:2022ocp}).
Therefore, our analysis in the GM-VFN scheme is supposed to enhance our previous result  in the ZM-VFNs \cite{Baradaran:2025khn} by retaining all non-logarithmic $m_b$-terms.\\
To take into account all influencing factors, we also include finite-$m_B$ effects, 
which modify the relations between hadronic and partonic scaling variables and reduce the available phase space. 
Their study is essential   to fully exploit the 
enormous statistics of the data from the LHC and future lepton colliders  for a high precision determination of the Higgs properties. Our analysis  shows that  the B-meson  mass is responsible for a significant enhancement of energy distribution  in the low-$x_B$ region while the bottom quark mass  leads  to an enhancement  in the peak region and above.  The masses of hadrons are also responsible for the low-$x_B$ threshold. 

The manuscript is organized as follows.
In Sec.~\ref{sec:two}, we briefly describe our applied framework to compute  the differential decay rates at the parton level and present our analytical results at NLO pQCD. In Sec.~\ref{sec:three} we review the factorization theorem in the presence of hadron mass. 
In Sec.~\ref{sec:four}, our numerical results at  hadron-level are presented and 
 Sec.~\ref{sec:five} is devoted to the conclusions.

\boldmath
\section{Parton level results for $H\to b\bar{b}$}
\label{sec:two}
\unboldmath

\boldmath
\subsection{Born level rate }
\unboldmath

In order to obtain the energy distribution of B-mesons inclusively produced in Higgs decay, i.e., $H\to  B+Jets$, we first need to compute  the differential  decay width of the process $H\to b\bar{b}$, i.e., $d\tilde\Gamma_i(H\to b\bar{b}(+g))/dx_i (i=b,g)$ at each order of perturbation. Here,  $x_i(=2p_i\cdot p_H/p_H^2)$ stands for the scaled-energy of b-quarks or gluon. In the Higgs rest frame, this  is simplified as $x_i=2E_i/m_H$. The gluon contribution appears in NLO and higher orders.\\
At the lowest-order approximation (Born level), the amplitude for the process $H\rightarrow b\bar{b}$ is parameterized
 as \textit{$M_0\propto (-im_b/v)v_{\bar{b}}\bar{u}_b$} where $v=(\sqrt{2}G_F)^{-1/2}$ and  $G_F$ is the Fermi constant.
In the Higgs boson rest frame, the tree-level decay width is given by
\begin{eqnarray}\label{gammatree}
\tilde\Gamma_0=\frac{N_c }{8\pi v^2}m_H^3 R\beta^3,
\end{eqnarray}
where, $R=(m_b/m_H)^2$, $\beta=\sqrt{1-4R}$ and $N_c=3$ is the color number of flavors.  This rate  is in complete agreement with the result presented in Refs.~\cite{Li:1990ag,Braaten:1980yq,Kataev:1993be,Sakai:1980fa}.\\
In the following, the technical detail of our computations  for the ${\cal O}(\alpha_s)$ QCD   radiative corrections to the Born level decay rate of $H\to b\bar{b}$ is explained using the dimensional regularization approach to regularize all divergences. For this calculation, we apply the fixed-flavor-number scheme (FFNs)  where $m_b\neq 0$ is assumed from the beginning.
 
\boldmath
\subsection{Virtual radiative corrections}
\label{virtual}
\unboldmath

\begin{figure}
	\begin{center}
		\includegraphics[width=0.470\textwidth]{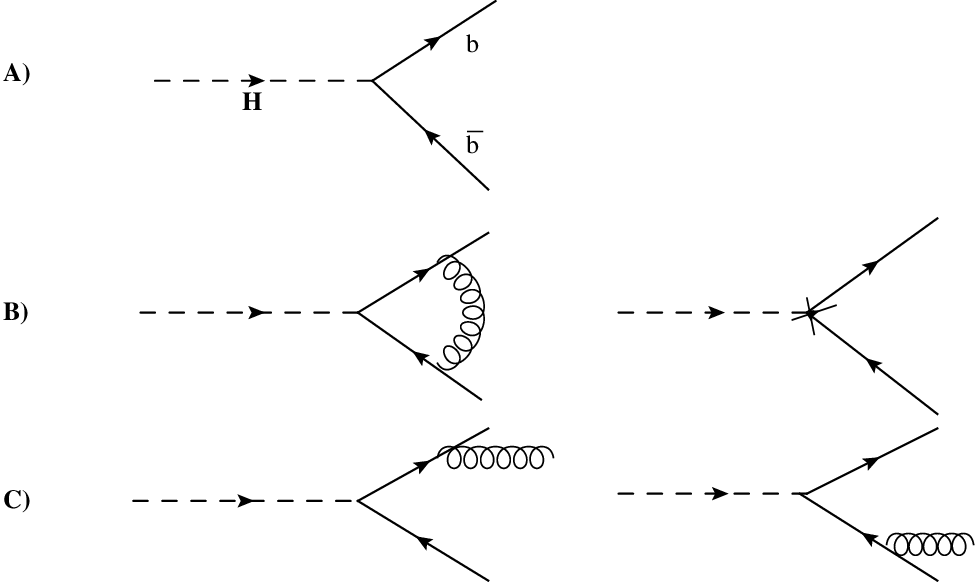}
		\caption{\label{plot}%
			Higgs decay Feynman diagrams at NLO: (A) tree-level contribution;  (B) vertex correction and a combination of mass and wave function renormalization; (C)   real radiations. 
 } 
	\end{center}
\end{figure}

Virtual radiative corrections to the $Hb\bar{b}$-vertex include  the one-loop vertex contribution and  a combination of b-quark mass and wave function renormalization, see Fig.~\ref{plot}B. These corrections   consist of both ultraviolet (uv) and infrared (ir) divergences which arise from the hard and soft virtual gluon radiations, respectively.  There  are different schemes for the renormalization of physical quantities. In this work we adopt
the on-shell mass-renormalization scheme and to regularize all singularities we apply the dimensional
regularization approach in which $D=4-2\epsilon$ is adopted for the dimension of space-time ($\epsilon\ll 1$).
 Then, all singularities appear as  $\epsilon_{uv}$ or $\epsilon_{ir}$.
To determine the energy spectrum of B-mesons produced in  $H\to b\bar{b}(+g)\to B+X$, where $X$ collectively denotes the unobserved final state particles,  we first need to have the partial decay widths $d\tilde\Gamma_i/dx_i (i=b, g)$ where the scaled energies $x_i$ in the Higgs rest frame are defined as before: $x_i=2E_i/m_H$.
The contribution of virtual gluon radiation  into the
differential decay width $d\tilde\Gamma_b/dx_b$ is computed through the following relation
\begin{eqnarray}
\frac{d\tilde\Gamma^{\textbf{vir}}_b}{dx_b}=\frac{\beta}{8\pi m_H}\overline{|M^{\textbf{vir}}|^2}
\delta(1-x_b).
\end{eqnarray} 
Here,
$\overline{|M^{\textbf{vir}}|^2}=N_c(m_b/v)^2\sum_{Spin}(M_0^{\dagger} M_{loop}+M_{loop}^{\dagger} M_0)$ where \textit{$M_0$} is the amplitude at the Born level, i.e, \textit{$M_0\propto (-im_b/v)v_{\bar{b}}\bar{u}_b$}. The  renormalized amplitude is written as  
\begin{eqnarray}
M_{loop}=v_{\bar{b}}(\Lambda_{ct}+\Lambda_l)\bar{u}_b,
\end{eqnarray}
where $\Lambda_{ct}$ stands for the counter-term and $\Lambda_l$  arises from the one-loop vertex correction, see Fig.~\ref{plot}B. The  counter-term of the vertex includes  a combination of the mass and the wave-function renormalization of $b/\bar{b}$-quarks which are expressed as \cite{Braaten:1980yq,kadeer, Czarnecki:1992ig, Liu:1992qd,MoosaviNejad:2012ju}
\begin{eqnarray}
\Lambda_{ct}&=& \delta Z_b+
\frac{\delta m_b}{m_b},
\end{eqnarray}
where, the renormalization constants are \cite{Corcella:1,MoosaviNejad:2011yp}
\begin{eqnarray}\label{mass}
\frac{\delta m_b}{m_b}&=&-\frac{\alpha_s(\mu_R)}{4\pi}C_F\bigg(\frac{3}{\epsilon_{uv}}-3\gamma_E+
3\ln\frac{4\pi\mu_F^2}{m_b^2}+4\bigg),
\nonumber\\
\delta Z_b &=&- \frac{\alpha_s(\mu_R)}{4\pi}C_F\bigg(\frac{1}{\epsilon_{uv}}+\frac{2}{\epsilon_{ir}}-3\gamma_E
\nonumber\\
&&\hspace{3cm}+3\ln\frac{4\pi\mu_F^2}{m_b^2}+4\bigg).
\end{eqnarray}
Here, $\gamma_E=0.577216\cdots$ is the Euler constant, $\alpha_s$ is the strong coupling constant and  $C_F=(N_c^2-1)/(2N_c)=4/3$ is the color factor. Moreover, $\mu_R$ and $\mu_F$ are the renormalization and factorization scales, respectively. For our analysis we adopt $\mu_R=\mu_F=\mu$, a choice often made. In addition, for the vertex correction in D-dimensions one has
\begin{eqnarray}
&&\Lambda_l=\frac{\pi\alpha_s C_F}{i(2\pi)^4}\times\\
&&\int \frac{d^Dp_g}{(2\pi\mu)^{D-4}}\gamma^\mu\frac{(\displaystyle{\not}\bar{p}_b+\displaystyle{\not}p_g-m_b)(\displaystyle{\not}p_b-\displaystyle{\not}p_g+m_b)}{p_g^2(p_b\cdot p_g)(\bar{p}_b\cdot p_g)}\gamma_\mu,\nonumber
\end{eqnarray}
where, $p_g, p_b$ and $\bar{p}_b$ are the gluon and $b/\bar{b}$-quark four-momenta, respectively. Then, the contribution of vertex correction in the mean squared amplitude ($\overline{|M^{\textbf{vir}}|^2}$) is proportional to the following term  
\begin{eqnarray}
&&\frac{\alpha_s m_H^2}{\pi}C_F\big[4R-1+2(1-2R)B_0(m_b^2,0,m_b^2)\nonumber\\
&&-M_H^2(1-6R+8R^2) C_0(m_b^2,m_b^2,m_H^2,m_b^2,0,m_b^2)\nonumber\\
&&-4RB_0(m_H^2,m_b^2,m_b^2)],
\end{eqnarray}
where,  $B_0$ and $C_0$ are the Passarino-Veltman 2-point and 3-point integrals \cite{Dittmaier:2003bc}. The one-loop virtual contribution is purely real as can be understood from an inspection of the one-loop Feynman diagrams. These diagrams do not accept any nonvanishing physical two-particle cut. \\
After summing all virtual radiative corrections up the uv-divergences are canceled but the ir-singularities are remaining which are labeled by $\epsilon$ from now on.
Ignoring more details, the virtual differential decay rate reads
\begin{eqnarray}\label{virt}
&&\frac{d\tilde\Gamma^{\textbf{vir}}_b}{dx_b}=\tilde\Gamma_0
\frac{\alpha_s(\mu)}{2\pi}C_F
\delta(1-x_b)\bigg\{-2-\frac{4R}{\beta}\ln x_s^2\nonumber\\
&&-\frac{2(1-2R)x_s}{R(1-x_s^2)}\bigg[2Li_2(1-x_s)-Li_2(1-x_s^2)-\frac{1}{8}\ln^2x_s^2\bigg]
\nonumber\\
&&-2\bigg[1-\frac{(1-2R)x_s}{2R(1-x_s^2)}\ln x_s^2\bigg]\bigg(\ln\frac{4\pi \mu^2}{m_b^2}-
\gamma_E+\frac{1}{\epsilon}\bigg)\bigg\},
\end{eqnarray}
where,     $Li_2(x)=-\int_0^x(dt/t)\ln(1-t)$ is the Spence function and we also defined $x_s=(\beta-1)/(\beta+1)$.

\boldmath
\subsection{Real gluon radiative corrections  (Bremsstrahlung)}\label{real}
\unboldmath

According to the well-known Bloch-Nordsieck theorem, all  singularities cancel exactly to all orders of perturbation theory, leaving a finite radiative correction of order $\alpha_s$. Then, the real gluon radiative corrections are also needed to remove remaining ir-singularities.
Considering two Feynman diagrams in Fig.~\ref{plot}C for the process $H(p_H)\to b(p_b)+\bar{b}(\bar{p}_b)+g(p_g)$,  the ${\cal O}(\alpha_s)$ real gluon emission  amplitude is written as 
\begin{eqnarray}\label{finfin}
M^{\textbf{real}}&=&\frac{-im_b}{v} g_sT^a \bar{u}(p_b, s_b)\big\{\frac{\gamma^\nu \displaystyle{\not}p_g+2p_b^\nu }{2p_b \cdot p_g}
\nonumber\\
&&-\frac{\gamma^\nu\displaystyle{\not}p_g+2\bar{p}_b^\nu}{2\bar{p}_b \cdot p_g}\big\}
v(\bar{p}_b, \bar{s}_b)\epsilon_{\nu}^{\star}(p_g,r),
\end{eqnarray}
where, the $T^a$-generators  are related to the Gell-Mann matrices. Also, $\epsilon(p_g,r)$ stands for the polarization vector of  real emitted gluon with the four-momentum $p_g$ and the spin $r$. 
Using the dimensional regularization scheme the differential decay rate of process $H\to b\bar{b}g$ is given by
\begin{eqnarray}\label{moozmooz}
d\tilde\Gamma^{\textbf{real}}=\frac{\mu^{2(4-D)}}{2 m_H}\overline{|M^{\textbf{real}}|^2}dPS(p_b, \bar{p}_b, p_g, p_H).
\end{eqnarray}
Here, the phase space element $dPS$ reads
\begin{eqnarray}\label{ahah}
\frac{d^{D-1}\vec{p}_b}{2E_b}\frac{d^{D-1}\vec{\bar{p_b}}}{2\bar{E}_b}\frac{d^{D-1}\vec{p}_g}{2E_g}
(2\pi)^{3-2D}\delta^D(p_H-\sum_{g,b,\bar{b}} p_f).
\end{eqnarray}
To evaluate the real differential decay rate $d\tilde\Gamma^{real}_b/dx_b$, we fix the b-quark momentum in Eq.~(\ref{moozmooz}) and integrates over 
the gluon energy ($E_g$). As before, we define $x_g=2E_g/m_H$ in the Higgs rest frame then it ranges as
\begin{eqnarray}
\frac{(1-x_b)(2-x_b-S)}{2(1+R-x_b)}\leq x_g \leq  \frac{(1-x_b)(2-x_b+S)}{2(1+R-x_b)}, 
\end{eqnarray}
where $S=\sqrt{x_b^2-4R}$. When integrating over the phase 
space of the process, in the rate $d\tilde\Gamma^{real}_b/dx_b$  terms of the form $(1-x_b)^{-1-2\epsilon}$ appear which are due to the radiation of  soft gluon. Note that, the limit of $E_g\to 0$ corresponds to the limit $x_b\to 1$. Therefore, for a massive b-quark where $x_b^{min}=\eta$ (with $\eta=2\sqrt{R}$), we apply the following prescription introduced in Ref.~\cite{Corcella:1}
 \begin{eqnarray}
\frac{(x_b-\eta)^{2\epsilon}}{(1-x_b)^{1+2\epsilon}}=-\frac{1}{2\epsilon}\delta(1-x_b)+
\frac{1}{(1-x_b)_+}+{\cal O}(\epsilon),
\end{eqnarray}
where  the plus description   is defined as 
\begin{eqnarray}  
\int_\eta^1 \frac{f(x_b)}{(1-x_b)_+}dx_b&=&\nonumber\\
&&\hspace{-2.5cm}\int_\eta^1\frac{f(x_b)-f(1)}{1-x_b}dx_b+f(1)\ln(1-\eta).
\end{eqnarray}

\boldmath
\subsection{Parton-level results of $d\tilde\Gamma_i/dx_i$ in FFN scheme}
\unboldmath

Ignoring more details, by summation all contributions, i.e., the tree-level, virtual and  real gluon radiative contributions, we  obtain an analytical result for the differential decay rate of $H\to b\bar{b}$ at NLO pQCD in the FFN scheme. The final result  is free of each singularity. Defining:
\begin{eqnarray}
F(x_b)=\ln\frac{x_b-2R+\sqrt{x_b^2-4R}}{x_b-2R-\sqrt{x_b^2-4R}},
\end{eqnarray}
the differential decay rate  normalized to the Born rate $\tilde{\Gamma}_0$ (\ref{gammatree}) reads:
\begin{eqnarray}\label{first}
&&\frac{1}{\tilde\Gamma_0}\frac{d\tilde\Gamma_b}{dx_b}=\delta(1-x_b)+
\frac{C_F\alpha_s(\mu)}{2\pi}(\frac{1-2 R}{\beta})\Bigg\{\nonumber\\
&&\delta(1-x_b)\bigg[\frac{4\beta}{1-2R}\ln\frac{R}{1-\eta}+\ln x_s^2\bigg(-\frac{1}{4}\ln x_s^2+\nonumber\\
&&\ln\frac{1+\eta}{1-\eta}-\frac{1+2R}{1-2R}\bigg)-2Li_2(1-x_s^2)+4Li_2(1-x_s)\bigg]+\nonumber\\
&&\frac{2\beta^{-2}(1-2 R)^{-1}}{(1-x_b)_+}\bigg[\bigg(8R^2-2R(2x_b+1)+x_b^2\bigg)F(x_b)+\nonumber\\
&&\frac{\sqrt{x_b^2-4R}}{2(1+R-x_b)}\bigg(16R^2-10Rx_b+6R+x_b^2-x_b\bigg)\bigg]+\nonumber\\
&&\frac{\beta^{-2}\sqrt{x_b^2-4R}}{2(1+R-x_b)(1-2R)}\bigg(11R-x_b-4+\frac{R(R-1)}{1+R-x_b}\bigg)\nonumber\\
&&+\frac{\beta^{-2}(1+x_b)}{1-2R}F(x_b)\Bigg\}.
\end{eqnarray}
This result, which is presented for the first time, after integrating over $x_b$ (with $2\sqrt{R}\le x_b\le 1$) is in complete agreement with the one presented in Refs.~\cite{Li:1990ag,Braaten:1980yq,Kataev:1993be,Sakai:1980fa}, i.e.,
\begin{eqnarray}\label{aida}
\tilde{\Gamma}(H\to b\bar{b})&=&\tilde{\Gamma}_0\Big\{1+\frac{C_F\alpha_s(\mu)}{\pi}\Big(\frac{G(\beta)}{\beta}-\nonumber\\
&&\hspace{-1.75cm}\frac{3(1-7\beta^2)}{8\beta^2}+\frac{3+34\beta^2-13\beta^4}{16\beta^3}\ln\frac{1+\beta}{1-\beta}\Big)\Big\},
\end{eqnarray}
where,
\begin{eqnarray}
G(\beta)&=&-3\beta\ln\frac{4}{1-\beta^2}-4\beta\ln\beta+\nonumber\\
&&(1+\beta^2)\Big(-2\ln\beta\ln\frac{1+\beta}{1-\beta}+3\ln (1+\beta)\ln\frac{1+\beta}{1-\beta}\nonumber\\
&&+4Li_2\frac{1-\beta}{1+\beta}+2Li_2\frac{\beta-1}{\beta+1}\Big).
\end{eqnarray}
Since, observed B-mesons from Higgs decays can  also be produced via the fragmentation of real gluons emitted at higher orders then, in order to have the most accurate energy spectrum of B-mesons at  NLO, we have to take into account  the contribution of gluon fragmentation too. Therefore, we also need to compute the differential decay rate $d\tilde\Gamma_g/dx_g$ in
the FFN scheme where $x_g(=2E_g/m_H)$  is the scaled-energy fraction of the real gluon emitted. We will show that the contribution of gluon splitting is important at the low-$x_B$  and decreases the size of partial decay width  at the threshold, see also Refs.~\cite{Kniehl:2012mn,MoosaviNejad:2016gcd,MoosaviNejad:2014uzv}. \\
In order to evaluate  the rate $d\tilde\Gamma_g/dx_g$, in Eq.~(\ref{moozmooz}) we integrate over the b-quark energy by  fixing the gluon  momentum in the phase space. The scaled energy of b-quark, i.e. $x_b=2E_b/m_H$, ranges as:
\begin{eqnarray}
1-\frac{x_g}{2}(1+v)\le x_b\le 1-\frac{x_g}{2}(1-v),
\end{eqnarray} 
where, $v=\sqrt{1-4R/(1-x_g)}$.\\
Ignoring more details, the   differential decay rate $d\tilde\Gamma_g/dx_g$ normalized to the Born rate $\tilde{\Gamma}_0$ (\ref{gammatree}) in the FFNs reads:
\begin{eqnarray}\label{second}
\frac{1}{\tilde\Gamma_0}\frac{d\tilde\Gamma_g}{dx_g}&=&\frac{C_F\alpha_s(\mu) }{2\pi \beta x_g}\bigg\{4(x_g-1)v\nonumber\\
&&+2\bigg(1+\beta^2+\frac{x_g^2}{\beta^2}-2x_g\bigg)\ln\frac{1+v}{1-v}\bigg\}.
\end{eqnarray}

As was explained, the GM-VFN scheme provides an ideal theoretical framework to study the effects of heavy quark masses. This scheme combines the virtues of the ZM-VFN and FFN schemes and avoids their flaws, at the same time.
Through this elaborate scheme the perturbative fragmentation functions enter the formalism via subtraction
terms for the hard scattering decay rates, so that the actual FFs are truly nonperturbative. They  may be assumed to have some smooth forms which can be specified through global data fits.
In contrast to the FFNs, the GM-VFNs  does also accommodate FFs for gluons and light flavor quarks, 
as in the ZM-VFN scheme. In our present study, the GM-VFNs is applied to resum the
large logarithms in $m_b$ and to retain the entire
nonlogarithmic $m_b$-dependence at the same time. This is reached by introducing suitable subtraction terms in the  ${\cal O}(\alpha_s)$ FFN expressions for $d\tilde\Gamma_i/dx_i$, so
that the ${\cal O}(\alpha_s)$ ZM-VFN results are exactly recovered in the limit $R\rightarrow 0(\equiv m_b/m_H\to 0)$.
These subtraction terms are universal quantities and so are the FFs in the FFN scheme, as is guaranteed by 
Collin's hard-scattering factorization theorem \cite{collins}.
The GM-VFN results for the decay distributions are obtained by matching the FFN results (\ref{first}, \ref{second})
to the ZM-VFN ones  by subtraction  as
\begin{eqnarray}\label{mahsa}
\frac{1}{\Gamma_0}\frac{d\Gamma_i}{dx_i}\Big|_{GM-VFN}=
\frac{1}{\tilde\Gamma_0}\frac{d\tilde\Gamma_i}{dx_i}\Big|_{FFN}-\frac{1}{\Gamma_0}\frac{d\Gamma_i}{dx_i}\Big|_{Sub},
\end{eqnarray}
where the subtraction terms are constructed as
\begin{eqnarray}
\frac{1}{\Gamma_0}\frac{d\Gamma_i}{dx_i}\Big|_{Sub}=
\lim_{m_b\rightarrow 0}\frac{1}{\tilde\Gamma_0}\frac{d\tilde\Gamma_i}{dx_i}\Big|_{FFN}-
\frac{1}{\hat\Gamma_0}\frac{d\hat\Gamma_i}{dx_i}\Big|_{_{ZM-VFN}}.
\end{eqnarray}
Analytical results for the terms $1/\hat\Gamma_0\times d\hat{\Gamma}^{NLO}/dx_i (i=b,g)$ at the ZM-VFN scheme  are given in our previous work \cite{Baradaran:2025khn}. Here, we just quote them as
\begin{eqnarray}\label{mar1}
&&\frac{1}{\hat{\Gamma}_0}\frac{d\hat\Gamma_b}{dx_b}\Big|_{_{ZM-VFN}}=\delta(1-x_b)+\nonumber\\
&&\hspace{0.8cm}\frac{C_F\alpha_s(\mu)}{2\pi}\Big\{\delta(1-x_b)\bigg(\frac{3}{2}+\frac{2\pi^2}{3}-\frac{3}{2}\ln\frac{\mu^2}{m_H^2}\bigg)-\nonumber\\
&&\hspace{0.75cm}\frac{1}{(1-x_b)_+}\big(\frac{3}{2}+(1+x_b^2)\ln\frac{\mu^2}{m_H^2}\big)+2\bigg(\frac{\ln(1-x_b)}{1-x_b}\bigg)_+\nonumber\\
&&\hspace{0.8cm}+\frac{5+x_b}{2}+2(\frac{1+x_b^2}{1-x_b})\ln x_b-(1+x_b)\ln(1-x_b)\Big\},\nonumber\\
\end{eqnarray}
and,
\begin{eqnarray}\label{mar2}
&&\frac{1}{\hat{\Gamma}_0}\frac{d\hat\Gamma_g}{dx_g}\Big|_{_{ZM-VFN}}=\nonumber\\
&&\hspace{1.2cm}\frac{C_F\alpha_s(\mu)}{2\pi}\Big\{\frac{2(1+(1-x_g)^2)}{x_g}\Big\}\Big(\frac{x_g^2}{1+(1-x_g)^2}\nonumber\\
&&\hspace{1.2cm}+\ln(1-x_g)+2\ln x_g-\ln\frac{\mu^2}{m_H^2}\Big),
\end{eqnarray}
where $\hat\Gamma_0=N_c m_H^3 R/(8\pi v^2)$ is the LO decay rate in the massless scheme. Taking the limit $m_b\rightarrow 0 (\equiv R\to 0)$ in Eqs.~(\ref{first}) and (\ref{second}), we recover Eqs.~(\ref{mar1}) and (\ref{mar2}) up to the terms
\begin{eqnarray}\label{ayda}
\frac{1}{\Gamma_0}\frac{d\Gamma}{dx_b}\Big|_{Sub}&=&\frac{\alpha_s(\mu)}{2\pi}C_F\Big\{\frac{1+x_b^2}{1-x_b}
\Big[\ln\frac{\mu^2}{m_b^2}\nonumber\\
&&-2\ln(1-x_b)-1\Big]\Big\}_+\, ,
\end{eqnarray}
and,
\begin{eqnarray}\label{mona}
\frac{1}{\Gamma_0}\frac{d\Gamma}{dx_g}\Big|_{Sub}&=&\frac{\alpha_s(\mu)}{2\pi}C_F\frac{2+2(1-x_g)^2}{x_g}\nonumber\\
&&\times\Big(\ln\frac{\mu^2}{m_b^2}-2\ln x_g-1\Big).
\end{eqnarray}
As is shown in Ref.~\cite{Kniehl:2012mn}, for the  top quark decay in the SM, i.e. $t\rightarrow bW^+$,
Eq.~(\ref{ayda}) coincides with the perturbative FF of the transition $b\rightarrow b$.
This is in consistency  with the Collin's factorization theorem which 
guarantees that the subtraction terms are universal, see also Refs.~\cite{Mitov:2004du,Melnikov:2004bm,Cacciari:2001cw,Ma:1997yq,Keller:1998tf,Yarahmadi:2022ocp}. 

\section{Hadron mass effects}
\label{sec:three}

Our main purpose is to evaluate  the scaled-energy ($x_B$) distribution 
of B-mesons  inclusively produced
in the SM Higgs decays at NLO.
Here, the  scaled-energy fraction of B-mesons is defined as $x_B=2p_B\cdot p_H/p_H^2$ which is simplified as $x_B=2E_B/m_H$ in the Higgs rest frame.\\
According to the Collin's factorization theorem of the QCD-improved parton model \cite{collins}, 
the energy distribution of B-mesons  is expressed as the convolution of hard scattering  decay rates at the parton level, i.e., $d\hat\Gamma_a/dx_a (a=b,g)$,  with the
nonperturbative FFs $D_{a=b,g}^B(z,\mu)$ which describe the transitions $(b,g)\rightarrow B$, i.e.,
\begin{eqnarray}\label{convolute}
&&\frac{d\Gamma}{dx_B}(H\to BX)=\nonumber\\
&&\sum_{a=b, g}\int_{x_a^{min}}^{x_a^{max}}
\frac{dx_a}{x_a}\frac{d\hat\Gamma_a}{dx_a}(x_a,\mu)D_a^B(z,\mu),
\end{eqnarray}
where $z=x_B/x_a\equiv E_B/E_a$ indicates the energy fraction of partons which is carrying away by the B-mesons. In Ref.~\cite{Kniehl:2012mn}, where the process of top quark decays for production of bottom-flavored mesons is studied, i.e., $t\to bW^+(+g)\to B+Jets$, authors have demonstrated that the above relation  is suitable for the massless cases (with $m_b=0$ and $m_B=0$). 
To calculate the quantity $d\Gamma/dx_B$ when
passing from the ZM-VFN scheme to the GM-VFN scheme
by taking into account the finite-$m_B$ corrections, one should apply the following improved factorization theorem 
\begin{eqnarray}\label{convolute2}
\frac{d\Gamma}{dx_B}(H\to BX)&=&\\
&&\hspace{-2.5 cm}\frac{1}{\sqrt{x_B^2-\rho_B^2}}\sum_{a=b, g}\int_{x_a^{min}}^{x_a^{max}}
dx_a z\frac{d\Gamma_a}{dx_a}\Big|_{_{GM-VFN}}D_a^B(z,\mu),\nonumber
\end{eqnarray}
where $(d\Gamma_a/dx_a)_{_{_{GM-VFN}}}$ is obtained via the approach defined in (\ref{mahsa}). Moreover, the energy fraction $z$ reads 
\begin{eqnarray}
z=\frac{x_B+\sqrt{x_B^2-\rho_B^2}}{x_a+\sqrt{x_a^2-\rho_a^2}},
\end{eqnarray}
where $\rho_i=m_i/E_b^{max}=2m_i/m_H(i=b, g, B)$.
Now, the kinematically allowed scaling-variables
are
\begin{eqnarray}
	&&\frac{1}{2}(x_B+\sqrt{x_B^2-\rho_B^2}+\frac{\rho_b^2}{x_B+\sqrt{x_B^2-\rho_B^2}})\leq x_b\leq 1,\nonumber\\
	&&\frac{1}{2}(x_B+\sqrt{x_B^2-\rho_B^2})\leq x_g\leq 1-4R,\nonumber
	\\
	&&\rho_B\leq x_B\leq 1-R+(\frac{m_B}{m_H})^2\quad \textrm{for b-quark transition},\nonumber
	\\
	&&\rho_B\leq x_B\leq 1-4R+(\frac{m_B}{m_H})^2\quad \textrm{for gluon transition}.\nonumber\\
	\end{eqnarray}
Clearly, if one sets $m_b=0$ and $m_B=0$ ($z\rightarrow x_B/x_a$), 
then Eqs.~(\ref{convolute}) and (\ref{convolute2}) coincide by reproducing the familiar
factorization formula of the massless parton model.

\boldmath
\section{numerical analysis}
\label{sec:four}
\unboldmath

To perform our phenomenological analysis  for the energy spectrum of B-mesons inclusively produced in Higgs decays
we adopt the following input parameters from Ref.~\cite{Nakamura:2010zzi}:
\begin{eqnarray}
&&G_F = 1.16637\times 10^{-5}~GeV^{-2},\nonumber\\
&&m_H = 125.3~GeV,\nonumber\\
&&m_b = 4.78~GeV,\nonumber\\
&&m_B = 5.279~GeV.
\end{eqnarray}
We evaluate the strong coupling constant at NLO  using the following relation \cite{Nakamura:2010zzi}:
\begin{eqnarray}\label{alpha}
\alpha^{(n_f)}_s(\mu^2)=\frac{1}{b_0\log(\mu^2/\Lambda^2)}
\Big\{1-\frac{b_1 \log\big[\log(\mu^2/\Lambda^2)\big]}{b_0^2\log(\mu^2/\Lambda^2)}\Big\},
\nonumber\\
\end{eqnarray}
where $n_f$ is the number of active quark flavors, and $b_0$ and $b_1$ are given by 
\begin{eqnarray}
b_0=\frac{33-2n_f}{12\pi}, \quad  b_1=\frac{153-19n_f}{24\pi^2}.
\end{eqnarray}
In Eq.~(\ref{alpha}),  $\Lambda$ is the  asymptotic scale parameter  for which we adopt
$\Lambda_{\overline{\text{MS}}}^{(5)}=231.0$~MeV adjusted such
that $\alpha_s^{(5)}=0.1184$ for $m_Z=91.1876$~GeV \cite{Nakamura:2010zzi}. Applying this equation one has $\alpha_s(m_H^2)$=0.113. 
\\
According to   the factorization theorems in the massless and massive schemes, i.e., Eqs.~(\ref{convolute},\ref{convolute2}),  we also  need the nonperturbative fragmentation functions $D_{a=b, g}^B(z, \mu)$ describing the hadronization processes $(b, g)\rightarrow B$. These functions indicate the   probability density of meson production from initial partons at the scale $\mu$. For these nonperturbative transitions, we use  the realistic nonperturbative FFs  determined at NLO through an overall  fit  to pair  annihilation data presented by ALEPH \cite{Heister:2001jg} and OPAL
\cite{Abbiendi:2002vt} at CERN LEP1 and by SLD \cite{Abe:1999ki} at SLAC SLC. Basically, in comparison with the hadron collisions, the $e^+e^-$ annihilation process provides a clean environment to determine the non-pFFs because, firstly the most precise data exist for these processes and secondly, one does  not deal with the uncertainties induced  by the parton distribution functions (PDFs).\\
In the literature, various phenomenological models have been offered to describe the hadronization mechanism. Among all models  the simple power model  as $D_b^B(z,\mu^\text{ini})=Nz^\alpha(1-z)^\beta$ is well-established. In Ref.~\cite{Salajegheh:2019ach},  authors have extracted the free parameters of this model, i.e., $N,\alpha$ and $\beta$, for  the fragmentation of all light and heavy flavors as well as gluon  into B-mesons at the initial scale $\mu^\text{ini}=4.5$~GeV. Their fit for the $b\to B$ splitting yielded $N=2575.014$, $\alpha=15.424$, and $\beta=2.394$. Furthermore, the gluon FF is generally assumed to be zero at the initial scale  and is generated
via the DGLAP  evolution equations \cite{dglap}
\begin{eqnarray}
&&\frac{d}{d\ln\mu^2}D_i^B(z,\mu)=\nonumber\\
&&\frac{\alpha_s(\mu)}{2\pi}\sum_j\int_z^1\frac{dy}{y}P_{ji}(y,\alpha_s(\mu))D_j^B(\frac{z}{y},\mu),
\end{eqnarray}
where $P_{ji}$ are the well-known splitting functions.

\begin{figure}
	\begin{center}
		\includegraphics[width=0.5\textwidth]{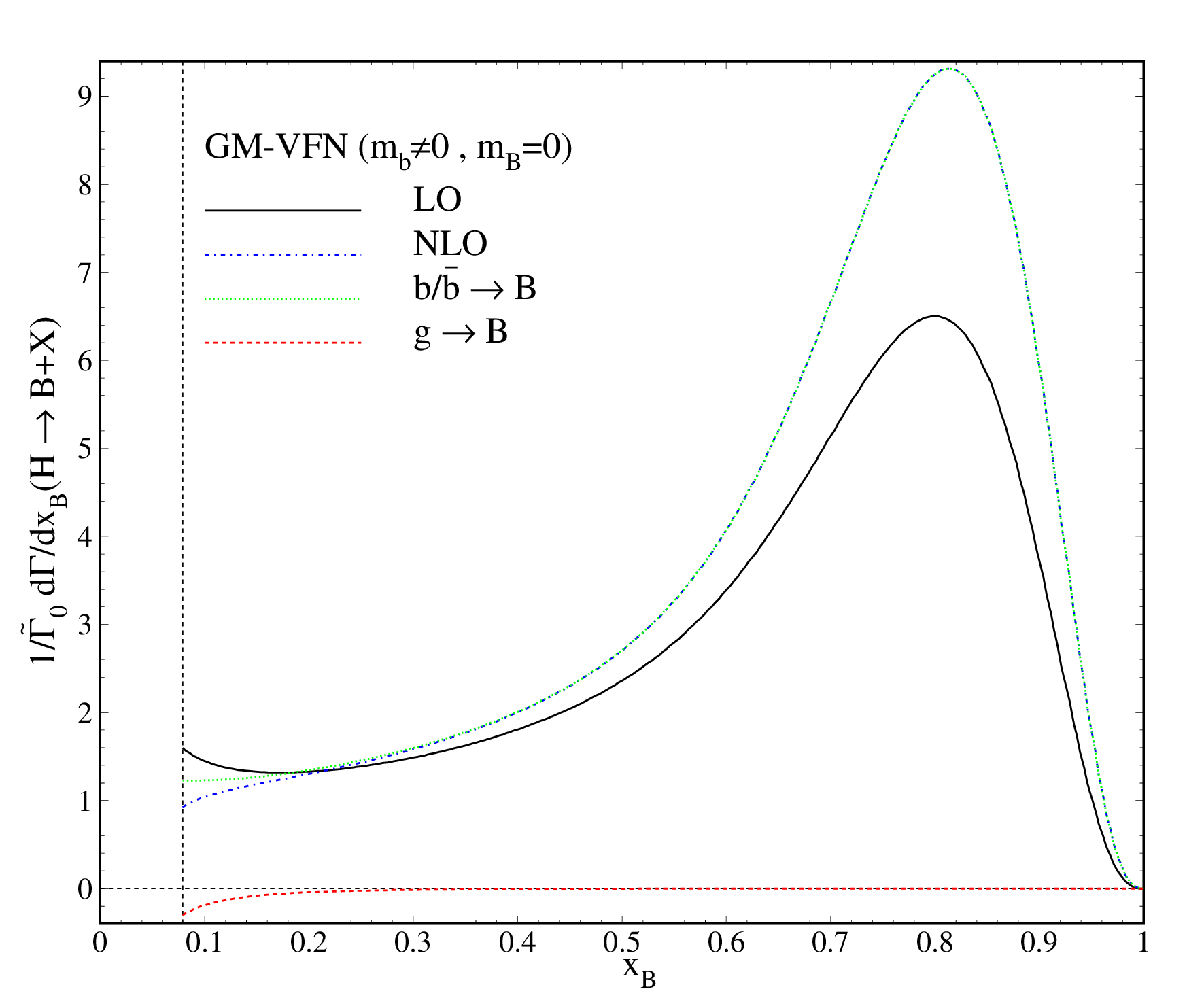}
		\caption{\label{fig1}%
			$1/\tilde\Gamma_0\times d\Gamma(H\to B +X)/dx_B$ as a function of $x_B$ in the GM-VFN  scheme. 
			The NLO result (blue dot-dashed line) is compared to the LO one (black solid line) and broken up into the contributions due to $b\to B$ (green dotted line) and $g\to B$ (red dashed line) fragmentation. Here we set $m_B=0$.
		} 
	\end{center}
\end{figure}

Assuming $\mu=m_H$ and $m_B=0$, in Fig.~\ref{fig1}  we study the NLO energy distribution of B-mesons in Higgs decays  using the GM-VFN scheme (with $m_b\neq 0$). For this aim, we consider  the normalized rate $(1/\tilde\Gamma_0)(d\Gamma/dx_B)$ where $\tilde\Gamma_0$ is given in Eq.~(\ref{gammatree}). For this prediction, we investigate  the size of NLO corrections by comparing the LO (black solid line) and NLO (blue dot-dashed line) results, and the relative importance of the $b\to B$ (green dotted line) and $g\to B$ (red dashed line) fragmentation channels at NLO. To expose the size of NLO radiative corrections at the parton level we evaluated the LO result using the same NLO FFs. We see that the NLO improvements  lead to an important  enhancement of partial decay width in the peak region ($x_B\approx 0.8$) and above by as much as $46\%$, at the expense of a reduction in the lower-$x_B$ range. 
Moreover, the peak position of energy spectrum is shifted towards higher amounts of $x_B$. It is also observed that the  contribution of $g\to B$ is throughout negative and appreciable only in the low-$x_B$ region ($x_B<0.3$). For higher-$x_B$ the NLO result is practically exhausted by the  contribution of $b\to B$, as we expect. Furthermore,  the finite-$m_b$ corrections are responsible for the appearance of a threshold.

\begin{figure}
	\begin{center}
		\includegraphics[width=0.5\textwidth]{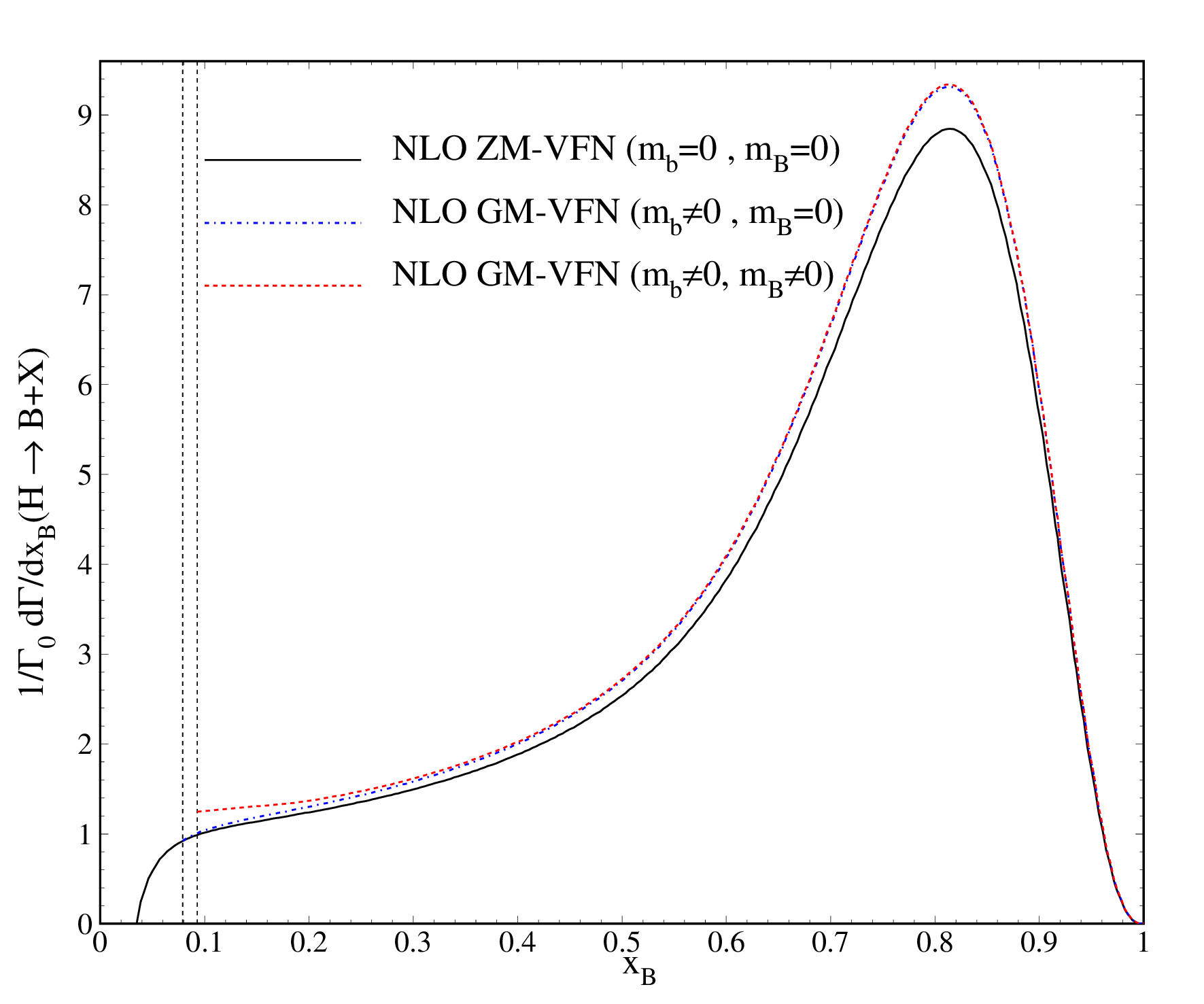}
		\caption{\label{fig2}%
			$1/\Gamma_0\times d\Gamma(H\to B +X)/dx_B$ as a function of $x_B$ at NLO. The GM-VFN ($m_b\neq 0$) results with (red dashed) and without finite-$m_B$ corrections (blue dot-dashed) are compared to the ZM-VFN ($m_b=0$) result for $m_B=0$ (black solid line). 
		} 
	\end{center}
\end{figure}

To study the mass effects of b-quark and B-meson, separately, in Fig.~\ref{fig2} we study the energy spectrum of B-mesons in three cases: a)- full massless or ZM-VFN scheme where one sets $m_b=0=m_B$ (black solid line), b)- massive or GM-VFN scheme (blue dot-dashed line) and c)- GM-VFN scheme with finite-$m_B$ corrections (red dashed curve).    It is observed that the finite-$m_B$ corrections are responsible for the appearance of a threshold at $x_B=0.093$. The finite mass corrections, generally, lead to an  enhancement in size throughout the whole $x_B$ range allowed and, specifically  the finite-$m_B$ corrections lead to the considerable enhancement in the region $0.093\le x_B\le 0.25$. Enhancements in other regions ($0.25\le x_B\le 1$) are due to the finite-$m_b$ correction which is more important than the effect of B-meson mass.  

\begin{figure}
	\begin{center}
		\includegraphics[width=0.5\textwidth]{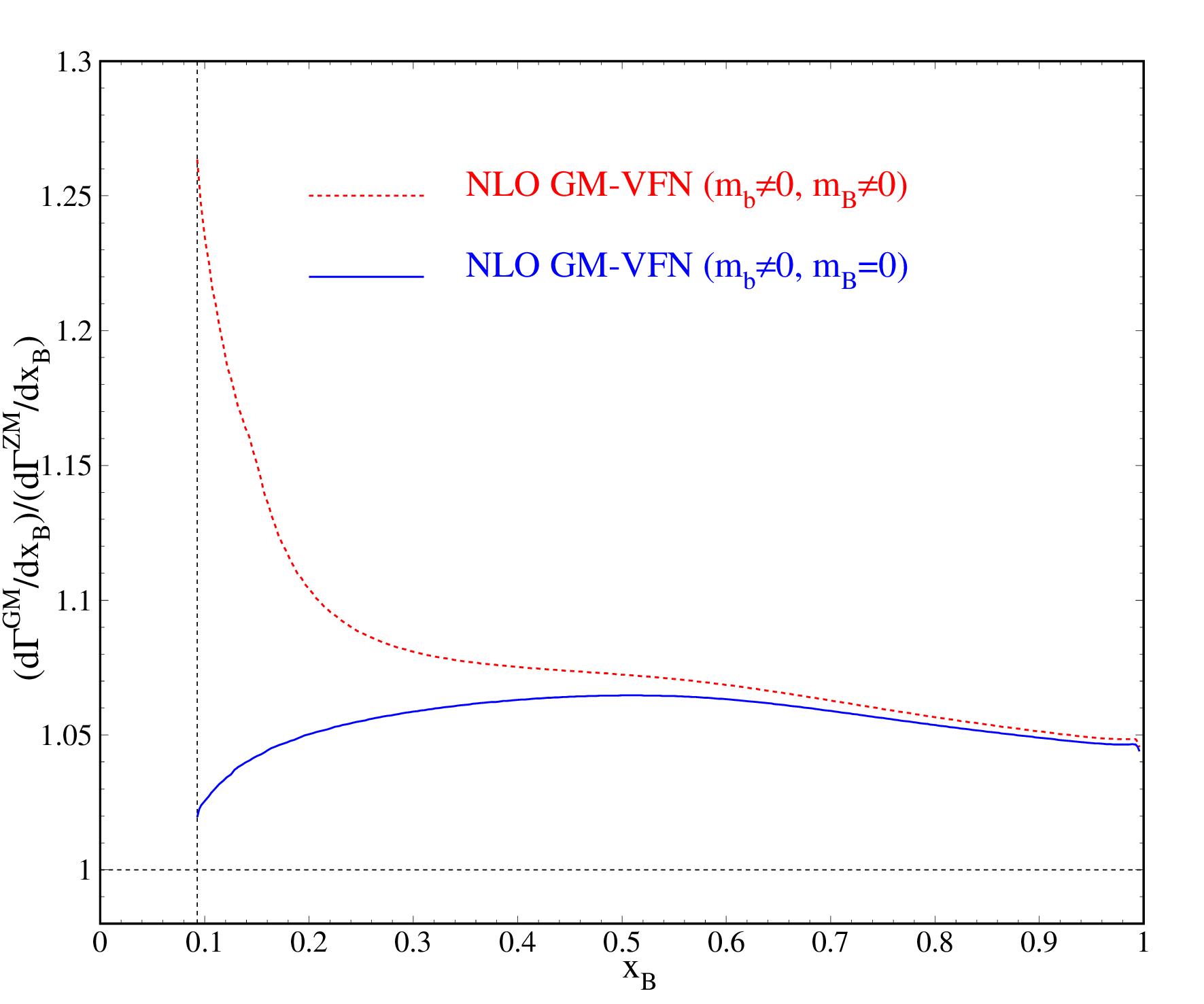}
		\caption{\label{fig3}%
			$1/\Gamma_0\times d\Gamma(H\to B +X)/dx_B$ as a function of $x_B$ at NLO in the GM-VFN ($m_b\neq 0$) scheme with (red dashed line) and without (blue solid line) finite-$m_B$ corrections. Both results are normalized to the ZM-VFN result  ($m_b=0$) for $m_B=0$. Here, we set $\mu=m_H$.		
		} 
	\end{center}
\end{figure}

For a more quantitative interpretation of Fig.~\ref{fig2}, we consider in Fig.~\ref{fig3} the GM-VFN results with (red dashed line) and without (blue solid line) finite-$m_B$ corrections  both normalized to the full ZM-VFN result for $m_B=0$. We observe that the inclusion of finite-$m_B$ and $m_b$  corrections lead to an enhancement of the partial decay width in size throughout the whole $x_B$ range allowed. Specifically, the finite-$m_B$ corrections leads to a significant enhancement of the partial decay width in the lower-$x_B$ range so that it increases the size of partial decay rate about $27\%$ at the threshold. For higher values of $x_B$, the $m_b$ effect is governed. For example,  the size of partial decay width increases  about $6\%$ at $x_B\approx 0.55$.

\begin{figure}
	\begin{center}
		\includegraphics[width=0.5\textwidth]{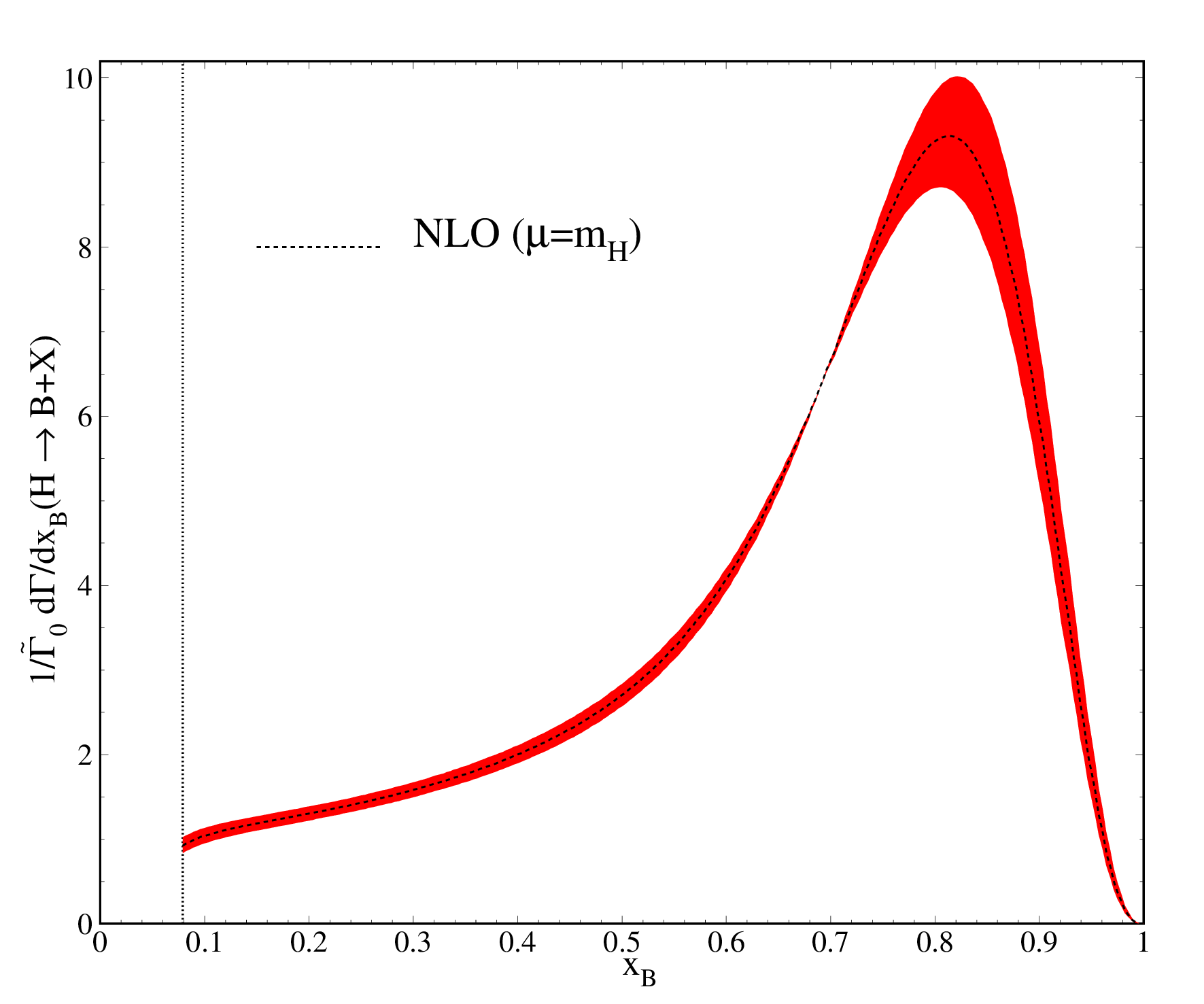}
		\caption{\label{fig4}%
			$1/\tilde\Gamma_0\times d\Gamma(H\to B +X)/dx_B$ as a function of $x_B$ at NLO in the GM-VFN scheme  considering various values for the scale $\mu$, i.e.  $m_H/2\le \mu\le 2m_H$. The result for $\mu=m_H$ is also shown (black dashed line). 
		} 
	\end{center}
\end{figure}

As was previously mentioned, in Eqs.~(\ref{convolute},\ref{convolute2}) we can use different values for the  scale $\mu$; however, in most literature to avoid large logarithms $\ln(\mu^2/m_H^2)$ in partonic differential decay rates a choice often made consists of setting $\mu=m_H$ (as we  assumed  in Figs.~\ref{fig1}-\ref{fig3}). Here, for studying the variation effect of the  scale $\mu$, in Fig.~\ref{fig4} we consider different choices for that as: $m_H/2\le \mu \le 2m_H$. For comparison, the result for $\mu=m_H$ is also shown (black dashed line).  As is seen, the effect of scale variation is considerable in the peak region and above.
\\
The formalism elaborated here is also usable for production of hadron species other than B-mesons, e.g. protons, pions, D-mesons, etc., through Higgs decay. The corresponding FFs $(b, g)\rightarrow \pi/P/D^+$  are given in Refs.~\cite{MoosaviNejad:2025leh,MoosaviNejad:2015lgp,MoosaviNejad:2017bda,Celiberto:2022dyf}.

\boldmath
\section{Summary and conclusions}
\label{sec:five}
\unboldmath

Since the discovery of Higgs boson in 2012 at the CERN-LHC, a lot of knowledge has been learned about this fundamental particle. Although, it is still required to  check  whether its characteristics, like couplings,  are completely consistent with those predicted by the SM so that any deviation from the SM predictions may be a signal of new effects. Its weak Yukawa coupling to the light fermions leaves space for theories of new physics beyond the SM and has attracted a lot of interest.
The high-luminosity LHC as well as the high-energy  LHC will allow physicists  to perform precise measurements that are impossible otherwise. This motivates particle physicists  to make a more detailed study of Higgs bosons, specifically via their  decay processes. 
Over $80\%$ of Higgs boson decays are fully hadronic, of which around $60\%$ decays into bottom quark pairs ($H\to b\bar{b}$). Bottom quarks hadronize before they decay and produce, in most cases, bottom-flavored hadrons (B-mesons). Therefore, in the present work we studied the process of B-mesons production in the decays of Higgs bosons at NLO pQCD, i.e., $H\to b\bar{b}(+g)\to B+Jets$.  For this study we evaluated  the distribution in the scaled-energy  of B-mesons which would be  of particular interest at the LHC and future colliders. In all previous works, for simplicity, the masses of bottom quarks as well as produced B-mesons have been ignored. In this work we, for the first time, preserved the bottom quark mass by working in the GM-VFN scheme  and computed an analytic
expression for  the NLO radiative corrections to the differential decay width of  Higgs bosons at the parton level, i.e., $d\tilde\Gamma_i^{NLO}(H\to \bar{b}b(+g))/dx_i\,\, (i=b, g)$ where $x_i=2E_i/m_H$ is the scaled-energy of partons in the Higgs rest frame. In the following,  using the nonperturbative FFs  of $(b, g)\to B$ the $x_B$-spectrum is predicted. 
We also analyzed the size of finite-$m_B$ effects on the energy spectrum as well as the theoretical uncertainty due to the freedom in the choice of factorization scale  $\mu$. Our results show that finite  masses are responsible for the low-$x_B$ threshold and a significant enhancement of the partial decay rate occurs in the low-$x_B$ due to $m_B$-finite mass while the b-quark mass  leads  to enhancements of the partial decay width in the peak region and above. The achievements of this research are suitable for more accurate studies of Higgs bosons in the high-energy and high-luminosity colliders such as the LHC and future colliders like FCC-ee and ILC.

\textbf{Data Availability Statement}: This manuscript has no associated
data. [Authors' comment: The needed input parameters can be found in the text and references of this paper.]

\textbf{ACKNOWLEDGMENTS} This work is based upon research funded by Iran National Science Foundation (INSF) under project NO.4039530.


\begin{thebibliography}{25}

\bibitem{CMS:2018zzl}
A.~M.~Sirunyan \textit{et al.} [CMS],
``Measurements of properties of the Higgs boson decaying to a W boson pair in pp collisions at $\sqrt{s}=$ 13 TeV,''
Phys. Lett. B \textbf{791} (2019), 96

\bibitem{CMS:2017zyp}
A.~M.~Sirunyan \textit{et al.} [CMS],
``Observation of the Higgs boson decay to a pair of $\tau$ leptons with the CMS detector,''
Phys. Lett. B \textbf{779} (2018), 283-316


\bibitem{CMS:2018nsn}
A.~M.~Sirunyan \textit{et al.} [CMS],
``Observation of Higgs boson decay to bottom quarks,''
Phys. Rev. Lett. \textbf{121} (2018) no.12, 121801


\bibitem{ATLAS:2018kot}
M.~Aaboud \textit{et al.} [ATLAS],
``Observation of $H \rightarrow b\bar{b}$ decays and $VH$ production with the ATLAS detector,''
Phys. Lett. B \textbf{786} (2018), 59-86

\bibitem{Cepeda:2019klc}
M.~Cepeda, S.~Gori, P.~Ilten, \textit{et al.}
``Report from Working Group 2: Higgs Physics at the HL-LHC and HE-LHC,''
CERN Yellow Rep. Monogr. \textbf{7} (2019), 221-584
[arXiv:1902.00134 [hep-ph]].


\bibitem{CEPCStudyGroup:2018ghi}
J.~B.~Guimar\~aes da Costa \textit{et al.} [CEPC Study Group],
``CEPC Conceptual Design Report: Volume 2 - Physics \& Detector,''
[arXiv:1811.10545 [hep-ex]].






\bibitem{FCC:2018evy}
A.~Abada \textit{et al.} [FCC],
``FCC-ee: The Lepton Collider: Future Circular Collider Conceptual Design Report Volume 2,''
Eur. Phys. J. ST \textbf{228} (2019) no.2, 261-623

\bibitem{ILC:2013jhg}
H.~Baer \textit{et al.} [ILC],
``The International Linear Collider Technical Design Report - Volume 2: Physics,''
[arXiv:1306.6352 [hep-ph]].

\bibitem{Black:2022cth}
K.~M.~Black, S.~Jindariani, D.~Li, F.~Maltoni, P.~Meade, D.~Stratakis, D.~Acosta, R.~Agarwal, K.~Agashe and C.~Aim\`e, \textit{et al.}
JINST \textbf{19} (2024) no.02, T02015.



\bibitem{Bodwin:2013gca}
G.~T.~Bodwin, F.~Petriello, S.~Stoynev and M.~Velasco,
``Higgs boson decays to quarkonia and the $H\bar{c}c$  coupling,''
Phys. Rev. D \textbf{88} (2013) no.5, 053003

\bibitem{Liao:2018nab}
Q.~L.~Liao, Y.~Deng, Y.~Yu, G.~C.~Wang and G.~Y.~Xie,
``Heavy $P$-wave quarkonium production via Higgs decays,''
Phys. Rev. D \textbf{98} (2018) no.3, 036014

\bibitem{Pan:2022nxc}
X.~A.~Pan, Z.~M.~Niu, M.~Song, Y.~Zhang, G.~Li and J.~Y.~Guo,
``J/\ensuremath{\psi} associated production with a bottom quark pair from the Higgs boson decay in next-to-leading order QCD,''
Phys. Rev. D \textbf{105} (2022) no.1, 014032

\bibitem{CDF:2015eag}
T.~A.~Aaltonen \textit{et al.} [CDF],
``Constraints on Models of the Higgs Boson with Exotic Spin and Parity using Decays to Bottom-Antibottom Quarks in the Full CDF Data Set,''
Phys. Rev. Lett. \textbf{114} (2015) no.14, 141802


\bibitem{Corcella:2004xv}
G.~Corcella,
``Fragmentation in H ---{\ensuremath{>}} b anti-b processes,''
Nucl. Phys. B \textbf{705} (2005), 363-383
[erratum: Nucl. Phys. B \textbf{713} (2005), 609-610]


\bibitem{Zheng:2023atb}
X.~C.~Zheng, X.~G.~Wu, X.~J.~Zhan, G.~Y.~Wang and H.~T.~Li,
``Higgs boson decays to Bc meson in the fragmentation-function approach,''
Phys. Rev. D \textbf{107} (2023) no.7, 074005





\bibitem{Baradaran:2025khn}
A.~Baradaran and S.~M.~Moosavi Nejad,
``Hadron production through Higgs boson decay at next-to-leading order in the zero-mass variable-flavor-number scheme,''
Eur. Phys. J. Plus \textbf{140} (2025) no.1, 41.

\bibitem{collins}
J.~C.~Collins,
``Finite-mass effects on inclusive $B$ meson hadroproduction,''
Phys.\ Rev.\  D {\bf 58}, 094002 (1998).

\bibitem{dglap}
V.~N.~Gribov and L.~N.~Lipatov,
``Deep Inelastic E P Scattering In Perturbation Theory,''
Sov.\ J.\ Nucl.\ Phys.\  {\bf 15}, 438 (1972)
[Yad.\ Fiz.\  {\bf 15}, 781 (1972)];
G.~Altarelli and G.~Parisi,
Nucl.\ Phys.\ {\bf B126}, 298 (1977);
Yu.~L.~Dokshitzer,
Sov.\ Phys.\ JETP {\bf 46}, 641 (1977)
[Zh.\ Eksp.\ Teor.\ Fiz.\  {\bf 73}, 1216 (1977)].




\bibitem{Kneesch:2007ey}
T.~Kneesch, B.~A.~Kniehl, G.~Kramer and I.~Schienbein,
``Charmed-meson fragmentation functions with finite-mass corrections,''
Nucl. Phys. B \textbf{799} (2008), 34-59.

\bibitem{Yarahmadi:2022ocp}
P.~S.~Yarahmadi and S.~M.~Moosavi Nejad,
``NLO corrections to the B-hadron energy distribution of heavy charged Higgs boson decay in the general-mass-variable-flavor-number scheme,''
Phys. Rev. D \textbf{106} (2022) no.5, 055040.

\bibitem{Li:1990ag}
C.~S.~Li and R.~J.~Oakes,
``QCD corrections to the hadronic decay width of a charged Higgs boson,''
Phys.\ Rev.\ D {\bf 43} (1991) 855.
doi:10.1103/PhysRevD.43.855


\bibitem{Kataev:1993be}
A.~L.~Kataev and V.~T.~Kim,
``The Effects of the QCD corrections to Gamma (H0 ---\ensuremath{>} b anti-b),''
Mod. Phys. Lett. A \textbf{9} (1994), 1309-1326


\bibitem{Sakai:1980fa}
N.~Sakai,
``Perturbative QCD Corrections to the Hadronic Decay Width of the Higgs Boson,''
Phys. Rev. D \textbf{22} (1980), 2220

\bibitem{Braaten:1980yq}
E.~Braaten and J.~P.~Leveille,
``Higgs Boson Decay and the Running Mass,''
Phys. Rev. D \textbf{22} (1980), 715


\bibitem{kadeer}
A.~Kadeer, J.~G.~K\"orner, and M.~C.~Mauser,
``A Phenomenological Study of Bottom Quark Fragmentation in Top Quark
Eur.\ Phys.\ J.\  C {\bf 54}, 175 (2008).


\bibitem{Czarnecki:1992ig}
A.~Czarnecki and S.~Davidson,
``On the QCD corrections to the charged Higgs decay of a heavy quark,''
Phys.\ Rev.\  D {\bf 47}, 3063 (1993).

\bibitem{Liu:1992qd}
J.~Liu and Y.~P.~Yao,
``QCD corrections to the charged Higgs boson decay of a heavy top quark,''
Phys.\ Rev.\  D {\bf 46}, 5196 (1992).

\bibitem{MoosaviNejad:2012ju}
S.~M.~Moosavi Nejad,
``${\cal O}(\alpha_s)$ corrections to the B-hadron energy distribution of the top decay in the Minimal Supersymmetric Standard Model considering GM-VFN scheme,''
Eur. Phys. J. C \textbf{72} (2012), 2224



\bibitem{Corcella:1}
G.~Corcella and A.~D.~Mitov,
``Bottom quark fragmentation in top quark decay,''
Nucl.\ Phys.\  B {\bf 623}, 247 (2002).

\bibitem{MoosaviNejad:2011yp}
S.~M.~Moosavi Nejad,
``B-mesons from top-quark decay in presence of the charged-Higgs boson in the Zero-Mass Variable-Flavor-Number Scheme,''
Phys.\ Rev.\ D {\bf 85}, 054010 (2012).  

\bibitem{Dittmaier:2003bc}
S.~Dittmaier,
``Separation of soft and collinear singularities from one loop N point integrals,''
Nucl.\ Phys.\ B {\bf 675}, 447 (2003).  


\bibitem{Kniehl:2012mn}
B.~A.~Kniehl, G.~Kramer and S.~M.~M.~Nejad,
``Bottom-flavored hadrons from top-quark decay at next-to-leading order in the general-mass variable-flavor-number scheme,''
Nucl.\ Phys.\  B {\bf 862}, 720  (2012)  

\bibitem{MoosaviNejad:2016gcd}
S.~M.~Moosavi Nejad and M.~Balali,
``Hadron energy spectrum in polarized top quark decays considering the effects of hadron and bottom quark masses,''
Eur. Phys. J. C \textbf{76} (2016) no.3, 173

\bibitem{MoosaviNejad:2014uzv}
S.~M.~Moosavi Nejad and M.~Balali,
``Angular analysis of polarized top quark decay into $B$-mesons in two different helicity systems,''
Phys. Rev. D \textbf{90} (2014) no.11, 114017
[erratum: Phys. Rev. D \textbf{93} (2016) no.11, 119904]

\bibitem{Mitov:2004du}
A.~Mitov,
``Perturbative heavy quark fragmentation function through $\mathcal{O}(\alpha^2_s)$: Gluon initiated contribution,''
Phys. Rev. D \textbf{71} (2005), 054021.

\bibitem{Melnikov:2004bm}
K.~Melnikov and A.~Mitov,
``Perturbative heavy quark fragmentation function through $\mathcal{O}(\alpha^2_s)$,''
Phys. Rev. D \textbf{70} (2004), 034027.

\bibitem{Cacciari:2001cw}
M.~Cacciari and S.~Catani,
``Soft gluon resummation for the fragmentation of light and heavy quarks at large x,''
Nucl. Phys. B \textbf{617} (2001), 253-290.

\bibitem{Ma:1997yq}
J.~P.~Ma,
``Perturbative prediction for parton fragmentation into heavy hadron,''
Nucl. Phys. B \textbf{506} (1997), 329-347.

\bibitem{Keller:1998tf}
S.~Keller and E.~Laenen,
``Next-to-leading order cross-sections for tagged reactions,''
Phys. Rev. D \textbf{59} (1999), 114004.





\bibitem{Nakamura:2010zzi}
K.~Nakamura {\it et al.}\  (Particle Data Group),
``Review of particle physics,''
J.\ Phys.\ G {\bf 37}, 075021 (2010).


\bibitem{Heister:2001jg}
A.~Heister {\it et al.}\  (ALEPH Collaboration),
``Study of the fragmentation of b quarks into B mesons at the Z peak,''
Phys.\ Lett.\  B {\bf 512}, 30 (2001).



\bibitem{Abbiendi:2002vt}
G.~Abbiendi {\it et al.}\  (OPAL Collaboration),
``Inclusive analysis of the b quark fragmentation function in Z decays at
LEP. ((B)),''
Eur.\ Phys.\ J.\  C {\bf 29}, 463 (2003).
[arXiv:hep-ex/0210031].


\bibitem{Abe:1999ki}
K.~Abe {\it et al.}\  (SLD Collaboration),
``Precise measurement of the b-quark fragmentation function in Z0 boson
Phys.\ Rev.\ Lett.\  {\bf 84}, 4300 (2000);
Phys.\ Rev.\  D {\bf 65}, 092006 (2002);
{\bf 66}, 079905 (2002).

\bibitem{Salajegheh:2019ach}
M.~Salajegheh, S.~M.~Moosavi Nejad, H.~Khanpour, B.~A.~Kniehl and M.~Soleymaninia,
``$B$-hadron fragmentation functions at next-to-next-to-leading order from a global analysis of $e^+e^-$ annihilation data,''
Phys.\ Rev.\ D {\bf 99} (2019) no.11,  114001.
doi:10.1103/PhysRevD.99.114001






\bibitem{MoosaviNejad:2015lgp}
S.~M.~Moosavi Nejad, M.~Soleymaninia and A.~Maktoubian,
``Proton fragmentation functions considering finite-mass corrections,''
Eur. Phys. J. A \textbf{52} (2016) no.10, 316


\bibitem{MoosaviNejad:2017bda}
S.~M.~Moosavi Nejad and M.~Delpasand,
``Polarized heavy baryon production in quark-diquark model considering two different scenarios,''
Eur. Phys. J. A \textbf{53} (2017) no.9, 174

\bibitem{MoosaviNejad:2025leh}
S.~M.~Moosavi Nejad,
``Heavy tetraquark production in high energy colliders through the fragmentation mechanism in a diquark model,''
Phys. Rev. D \textbf{112} (2025) no.5, 056020

\bibitem{Celiberto:2022dyf}
F.~G.~Celiberto and M.~Fucilla,
``Diffractive semi-hard production of a $J/\psi $ or a $\Upsilon $ from single-parton fragmentation plus a jet in hybrid factorization,''
Eur. Phys. J. C \textbf{82} (2022) no.10, 929

\end{thebibliography}
\end{document}